\documentclass[12pt]{hal}
\usepackage{amsfonts}

\newcommand{\sch}{Schr\"{o}dinger\ } 

\newcommand{\JCPformat}[4]{{#1} {\bf #2}, {#3} ({#4}).}

\newcommand{\Ref}[4]{\JCPformat{#1}{#2}{#3}{#4}}

\newcommand{\jcp}[3]{\Ref{J. Chem. Phys}{#1}{#2}{#3}}

\newcommand{\cp}[3]{\Ref{Chem. Phys.}{#1}{#2}{#3}}

\newcommand{\tca}[3]{\Ref{Theor. Chim. Acta.}{#1}{#2}{#3}}
\newcommand{\jcompchem}[3]{\Ref{J. Comp. Chem.}{#1}{#2}{#3}}
\newcommand{\jmathchem}[3]{\Ref{J. Math. Chem.}{#1}{#2}{#3}}
\newcommand{\jmathphys}[3]{\Ref{J. Math. Phys.}{#1}{#2}{#3}}

\newcommand{\ijqc}[3]{\Ref{Int. J. Quantum Chem.}{#1}{#2}{#3}}

\newcommand{\physrev}[3]{\Ref{Phys. Rev.}{#1}{#2}{#3}}

\newcommand{\jphys}[3]{\Ref{J. Phys.}{#1}{#2}{#3}}

\begin{document}
\begin{frontmatter}

\title{Geometric Measure of Indistinguishability for Groups of Identical Particles}
\author{Patrick Cassam-Chena\"{\i}}
\address{
 Laboratoire J. A. Dieudonn\'e, UMR 6621 du CNRS, Facult\'e des Sciences, Parc Valrose,
  06108 Nice cedex 2, France. cassam@unice.fr}

\date{}
\maketitle
\begin{abstract}
The concept of $p$-orthogonality ($1\leq p\leq n$) between $n$-particle states is introduced.
It generalizes common orthogonality, which is equivalent to $n$-orthogonality, and strong orthogonality between fermionic states, which is equivalent to $1$-orthogonality. Within the class of non $p$-orthogonal states a finer measure of non $p$-orthogonality is provided by Araki's angles between $p$-internal spaces. The $p$-orthogonality concept is a geometric measure of indistinguishability that is independent of the representation chosen for the quantum states. It induces a new hierarchy of approximations for group function methods. The simplifications that occur in the calculation of matrix elements between $p$-orthogonal group functions are presented. 

\end{abstract}
\end{frontmatter}
\newpage
\section{Introduction}

Recent studies have aimed at defining a geometric measure of entanglement (see \cite{Cao07} and therein).
Entanglement is also related to the Von Neumann entropy of reduced density operators, however, in the case of quantum systems made of identical particles, it has proved important to take apart the uncertainty due to the indistinguishability of the identical particles from that due to entanglement \cite{Ghirardi04}. In this work we provide a geometric measure of indistinguishability.

Indistinguishability of identical particles is related to orthogonality properties of Hilbert subspaces. It is common knowledge that sets of identical fermions can be considered as distinguishable, when their respective wave functions (or density operators) are built from one-particle functions belonging to orthogonal Hilbert spaces  \cite{Blaizot86}. In other words, when any one-particle state of a set of particles does not overlap with any one-particle state of another set, then antisymmetrizing or not antisymmetrizing the tensor product of the wave functions of the two sets give the same physical predictions. Such sets of fermions are said strongly orthogonal to each other \cite{Parr56,McWeeny59}.
However, as far as we are aware, when some one-particle states of the two sets do have non-zero overlap, so that the sets become indistinguishable, there is no measure to quantify to which extend the particles of both sets are actually mixed.

More specifically, let, $a^{\dagger}_1,\ldots ,a^{\dagger}_{2n}$ be $2n$ creation operators of orthonormal one-particle (either boson or fermion) states. The $n$-particle states, $a^{\dagger}_1\cdots a^{\dagger}_{n}\vert0\rangle$, and, $a^{\dagger}_1 \cdots  a^{\dagger}_{n-1} a^{\dagger}_{n+1}\vert0\rangle$, are orthogonal. So are, $a^{\dagger}_1 \cdots  a^{\dagger}_{n}\vert0\rangle$, and, $a^{\dagger}_{n+1} \cdots  a^{\dagger}_{2n}\vert0\rangle$.  
Intuitively the latter pair is ``more'' orthogonal than the former. 
In fact, it is ``strongly'' orthogonal \cite{Parr56,McWeeny59}. 
Between these two extreme cases, there are intermediate cases, like for example the pairs,  $a^{\dagger}_1 \cdots  a^{\dagger}_{n}\vert0\rangle$, and, $a^{\dagger}_{1} \cdots  a^{\dagger}_{n-p} a^{\dagger}_{n+1}\cdots  a^{\dagger}_{n+p}\vert0\rangle$, which are orthogonal but not strongly orthogonal.
The aim of the present article is to introduce a graded orthogonality concept which discriminates between all these cases. Our geometric concept is well-defined (i.e. independent of the arbitrarily chosen representation of quantum states) for general, multiconfigurational wave functions of possibly different particle numbers as well as for mixed (i.e ensemble) states. 

The article is organised as follow, we first recall the definition of the $p$-internal space of an $n$-particle quantum state, then we define the concept of $p$-orthogonality and the Araki's angles between the $p$-internal spaces, finally we show the usefulness of these concepts to simplify the calculation of matrix elements appearing in a class of general approximation methods for solving the \sch equation of $n$ identical particles. Note that throughout the article, the emphasis will be put on fermions, because $p$-orthogonality will be \textit{a priori} more useful for sets of particles obeying the Pauli principle, than for bosonic particles, whose states tend to degeneracy rather than to orthogonality.

\section{$p$-internal space of an $n$-particle state}

Let $\mathcal{H}$ denotes the one-particle Hilbert space and $\wedge^n\mathcal{H}$, (respectively, $\vee^n\mathcal{H}$), the Hilbert space of antisymmetric, (respectively, symmetric), $n$-particle states built from $\mathcal{H}$. Let $\Psi\in\wedge^n\mathcal{H}$ (respectively, $\Psi\in\vee^n\mathcal{H}$) be a normalized $n$-fermion (respectively, $n$-boson) wave function.
Its reduced density operator, $D_{\Psi }$ 
acts on a wave function, $\Phi\in\wedge\mathcal{H}$ (respectively, $\Phi\in\vee\mathcal{H}$) in the following way \cite{Cassam03-jmp},
\begin{equation}
D_{\Psi}(\Phi )=\Psi\hookrightarrow\Phi\hookleftarrow\Psi,
\label{redens} 
\end{equation} 
where $\hookrightarrow$ (resp. $\hookleftarrow$) denotes the right (resp. left)
interior product.

We recall that the interior products for fermions are defined by conjugation with respect to the Grassmann product: $\Theta\in\wedge^{q-p}\mathcal{H}$, 
$\Psi\in\wedge^p\mathcal{H}$, $\Phi\in\wedge^q\mathcal{H}$,
\begin{equation}
\langle\Theta |\Psi\hookleftarrow\Phi \rangle=\langle\Psi\wedge \Theta |\Phi \rangle ,
\label{int-l}
\end{equation}
\begin{equation}
\langle\Theta |\Phi \hookrightarrow\Psi\rangle=\langle\Theta\wedge\Psi |\Phi \rangle .
\label{int-r}
\end{equation}
Similarly, for bosonic states the interior products are conjugated to the symmetrical product $\vee$.

The interior products are equivalent to ``annihilation'' in the second quantization language; using this formalism  $D_{\Psi}(\Phi)$ would be written: $D_{\Psi}\vert\Phi\rangle=\sum\limits_{\Theta} \langle 0\vert\Psi\Theta^{\dagger}\Phi\Psi^{\dagger}\vert 0\rangle\ \vert\Theta\rangle$, where $\mathbf{\Psi, \Theta},$ (bold symbol) denote the annihilation  operators associated to $\Psi, \Theta$; $\mathbf{\Psi^{\dagger}, \Theta^{\dagger}},$ denote their conjugate creation operators. The reduced density operator preserves the number of particles, that is to say, $\wedge^p\mathcal{H}$, (respectively, $\vee^p\mathcal{H}$), is stable under $D_{\Psi}$.
The restriction of $D_{\Psi }$ to the $p$-particle space, $D^p_{\Psi }$, is the so called ``$p$-order reduced density operators'', (the action of $D^p_{\Psi }$ can be extended to the whole of $\wedge\mathcal{H}$,  (respectively, $\vee\mathcal{H}$) by $D^p_{\Psi }(\Phi)=0$ if $\Phi\in\wedge^q\mathcal{H}$ (respectively, $\vee^q\mathcal{H}$) and $q\neq p$, then $D_{\Psi }$ decomposes as a direct sum $D_{\Psi }=\sum_{p\geq 0} \ D^p_{\Psi }$).

We call ``$p$-internal space'' the sum of the eigenspaces of the $p$-order reduced density operator $D^p_{\Psi}$ associated to non zero eigenvalues. The $p$-particle functions of this space are called ``$p$-internal functions''.
An alternative definition of the $p$-internal space, of $\Psi\in\wedge^n\mathcal{H}$, denoted $\mathcal{I}^p[\Psi ]$, is:

\begin{equation}
\mathcal{I}^p[\Psi ]:=\{\Phi\in \wedge^{p}\mathcal{H}, \exists \Omega\in\wedge^{n-p}\mathcal{H},\Omega
\hookleftarrow\Psi =\Phi\} ,
\label{intspace-p}
\end{equation}
that is to say, $\mathcal{I}^p[\Psi ]$ is the vector space obtained by annihilating a $(n-p)$-fermion function in $\Psi$ in all possible manners. A similar definition holds for bosons.

Examples from electronic structure theory: The $1$-internal space, or simply ``internal space'', $\mathcal{I}^1[\Psi ]$, is the space spanned by the, so-called, occupied, natural spinorbitals, in quantum chemistry. The $2$-internal space, $\mathcal{I}^2[\Psi ]$, is the space spanned by the occupied, natural geminals. The $n$-internal space is the one-dimensional vector space spanned by the wave function $\Psi$. The $p$-internal space of a single configuration function (Slater determinant) built over a set of $n$ orthogonal spinorbitals, $\Psi:=\phi_1\wedge\ldots \wedge\phi_{n}$, is the ${n \choose p}$-dimensional vector space spanned by the $p$-particle functions, $\phi_{i_1}\wedge\ldots \wedge\phi_{i_p}$, built over $p$ spinorbitals of $\Psi$.

The definition extends to ensemble states described by a general density operators, $D^n$, that is, by a convex combination of pure states density operators:
\begin{equation}
D^n:=\sum_i\ c_i\ D^n_{\Psi_i }\qquad with \qquad c_i > 0\qquad and\qquad \sum_i\ c_i = 1.
\label{mixed_state_op}
\end{equation}
In such a case, the $p$-order reduced density operator to consider is, simply,
\begin{equation}
D^p=\sum_i\ c_i\ D^p_{\Psi_i }.
\label{R-mixed_state_op}
\end{equation}
It is easy to see that the $p$-internal space of $D^n$ is the sum (not necessarily direct) of its pure states $p$-internal space,
\begin{equation}
\mathcal{I}^p[D^n ]=\sum\limits_{i} \mathcal{I}^p[\Psi_i ].
\label{ens-intspace-p}
\end{equation}
The orthogonal complement of the $p$-internal space, that is the kernel of $D^p$, is called the $p$-external space, $\mathcal{E}^p[D^n ]:=\mathcal{I}^p[D^n ]^{\perp}$, and satisfies,
\begin{equation}
\mathcal{E}^p[D^n ]=\bigcap\limits_{i} \mathcal{E}^p[\Psi_i ].
\label{ens-extspace-p}
\end{equation}

\section{$p$-orthogonality}
\subsection{Definition}

Let $\Psi_1\in\wedge^{n_1}\mathcal{H}$ and $\Psi_2\in\wedge^{n_2}\mathcal{H}$ be respectively a  $n_1$- and a $n_2$-fermion wave function.

We will say that $\Psi_1$ and $\Psi_2$ are $p$-orthogonal (for $1\leq p\leq inf(n_1,n_2)$) if and only if their $p$-internal spaces are orthogonal,
\begin{equation}
\mathcal{I}^p[\Psi_1 ]\perp\mathcal{I}^p[\Psi_2 ].
\label{p-orthogonality}
\end{equation}
A similar definition holds for bosonic states, and extends to ensemble states, either bosonic or fermionic, by considering the orthogonality of the $p$-internal space of their associated density operators, $D_1^{n_1}$ and $D_2^{n_2}$.

We see immediately that if $n_1=n_2=n$, $n$-orthogonality is the usual orthogonality between wave functions. 
In the case of ensemble states, it means that any wave function associated to a pure state in the convex combination of one density operator is orthogonal to any wave function associated to a pure state in the convex combination of the other density operator.

At the other end, $1$-orthogonality between $\Psi_1$ and $\Psi_2$ amounts to strong orthogonality, usually defined by,
\begin{equation}
\int d\tau_1 \Psi_1(\tau_1,\tau_2,\ldots,\tau_{n_1})\Psi_2(\tau_1,\tau_2',\ldots,\tau_{n_2}') =0 \qquad\forall \tau_2,\ldots,\tau_{n_1},\tau_2',\ldots,\tau_{n_2}'.
\label{strong-orthogonality1}
\end{equation}
This can be rewritten, using Dirac distributions centered on the Fermion variables, as the nullity of the kernel,
\begin{equation}
\langle\delta_{\tau_2}\wedge\ldots\wedge\delta_{\tau_{n_1}}\hookleftarrow\Psi_1\vert\delta_{\tau_2'}\wedge\ldots\wedge\delta_{\tau_{n_2}'}\hookleftarrow\Psi_2\rangle =0 .
\label{strong-orthogonality2}
\end{equation}
or, by changing to a basis set representation $\{\phi_i\}_i$ in the rigged Hilbert space \cite{Carfi03}, as, 
\begin{equation}
\langle\phi_{i_2}\wedge\ldots\wedge\phi_{i_{n_1}}\hookleftarrow\Psi_1\vert\phi_{i_2'}\wedge\ldots\wedge\phi_{i_{n_2}'}\hookleftarrow\Psi_2\rangle=0\qquad\forall i_2,\ldots,i_{n_1},i_2',\ldots,i_{n_2}'.
\label{strong-orthogonality3}
\end{equation}
Since, the $(n_1-1)$-particle functions, $\phi_{i_2}\wedge\ldots\wedge\phi_{i_{n_1}}$, span all of, $\wedge^{(n_1-1)}\mathcal{H}$, and the $(n_2-1)$-particle functions, $\phi_{i_2'}\wedge\ldots\wedge\phi_{i_{n_2}'}$, span all of, $\wedge^{(n_2-1)}\mathcal{H}$, the latter equation is equivalent to  orthogonality between any pair of $1$-internal functions, that is to say to
$1$-orthogonality.

Remark: The present definition of ``strong orthogonality'' as the orthogonality of the one-internal spaces, and, another characterization of the one-internal space of a function $\Psi\in\wedge^n\mathcal{H}$  as the smallest Hilbert space, $\mathcal{F}$ such that $\Psi\in\wedge^n\mathcal{F}$ \cite{Cassam94}, make obvious the separability property of strongly orthogonal electron pairs \cite{Arai70}, or more generally, of strongly orthogonal electron groups (see also the definition of ``non-overlapping subsystems in \cite{Cassam04-pla}).
 
\subsection{Graded orthogonality}

An important property to notice is that $p$-orthogonality implies $q$-orthogonality for all $q\geq p$. Its proof relies essentially upon the following lemna:

\textbf{Lemna 1:} If $\Phi\in\mathcal{I}^{p+1}[D^n ]$ then $\forall \phi\in\mathcal{H}$, $(\phi
\hookleftarrow\Phi)\in\mathcal{I}^{p}[D^n ]$.\\
\textbf{Proof:} Consider first the case of fermionic pure states. Let $\Psi\in\wedge^n\mathcal{H}$ be an $n$-particle wave function, $\Phi\in\mathcal{I}^{p+1}[\Psi ]$ and $\phi\in\mathcal{H}$.  By Eq. (\ref{intspace-p}), there exists $\Omega\in\wedge^{n-p-1}\mathcal{H}$ such that $\Omega\hookleftarrow\Psi =\Phi$. So, $\phi\hookleftarrow\Phi=\phi\hookleftarrow\left( \Omega\hookleftarrow\Psi\right)  
=\left( \Omega\wedge\phi \right)\hookleftarrow\Psi$, where $\left( \Omega\wedge\phi\right) \in\wedge^{n-p}\mathcal{H}$. This means that $(\phi
\hookleftarrow\Phi)\in\mathcal{I}^{p}[\Psi ]$ according to Eq. (\ref{intspace-p}), and proves the proposition for fermionic pure states. The demonstration is the same for bosonic pure states with $\vee$ instead of $\wedge$.\\
Now consider a mixed state operator $D^n$ as in Eq. (\ref{mixed_state_op}), $\Phi\in\mathcal{I}^{p+1}[D^n ]$ and $\phi\in\mathcal{H}$. By Eq. (\ref{ens-intspace-p}), there exist $\Phi_i$'s such that $\Phi=\sum\limits_{i} \Phi_i$ and $\Phi_i\in\mathcal{I}^{p+1}[\Psi_i ]$ for all $i$. By (anti)linearity of the interior product, $\phi\hookleftarrow\Phi=\phi\hookleftarrow\left( \sum\limits_{i} \Phi_i\right)  
=\sum\limits_{i}\left( \phi\hookleftarrow\Phi_i \right)$. But we have just shown that $\left( \phi\hookleftarrow\Phi_i \right)\in\mathcal{I}^{p}[\Psi_i ]$ for all $i$, which proves the property for mixed states according to Eq. (\ref{ens-intspace-p}).

$p$-orthogonality is a graded property in the sense that:\\
\textbf{Proposition 1:} If two states represented by the density operators $D_1^{n_1}$ and $D_2^{n_2}$, (or by the wave functions $\Psi_1$ and $\Psi_2$ for pure states, with $D_i^{n_i}= \vert\Psi_i\rangle\langle\Psi_i\vert$), are $p$-orthogonal then they are \textit{a fortiori} $q$-orthogonal for all $q$ such that, $inf(n_1,n_2)\geq q \geq p$.\\  
\textbf{Proof:}\\
Let $1\leq p < n_1 \leq n_2$ be three integers.  Let $\{\phi_i\}_i$ be an orthonormal basis set of $\mathcal{H}$.  Consider the fermionic case, and note that the ``particle number'' operator, $\hat{N}:=\sum_i \phi_i\wedge\left( \phi_i\hookleftarrow \bullet\right) $, acts on the $q$-fermion space, $\wedge^{q}\mathcal{H}$, as, $q.Id_{\wedge^{q}\mathcal{H}}$, ($Id_{\wedge^{q}\mathcal{H}}$ denotes the identity on $\wedge^{q}\mathcal{H}$). 
For all  $\Gamma_1\in\mathcal{I}^{p+1}[D_1^{n_1} ]$, $\Gamma_2\in\mathcal{I}^{p+1}[D_2^{n_2} ]$, and $\phi_i\in\mathcal{H}$, we have $\langle \phi_i \hookleftarrow \Gamma_1 \vert\phi_i \hookleftarrow \Gamma_2\rangle=0$ since by lemna 1, $\left( \phi_i \hookleftarrow \Gamma_1\right) \in\mathcal{I}^{p}[D_1^{n_1} ]$, $\left( \phi_i \hookleftarrow \Gamma_2\right) \in\mathcal{I}^{p}[D_2^{n_2} ]$, and by hypothesis $D_1^{n_1}$ and $D_2^{n_2}$ are $p$-orthogonal. Therefore,
$0=\sum_i \langle \phi_i \hookleftarrow \Gamma_1 \vert\phi_i \hookleftarrow \Gamma_2\rangle
=\sum_i\langle \phi_i \wedge\left(\phi_i \hookleftarrow \Gamma_1\right) \vert \Gamma_2\rangle
= \langle \hat{N}\Gamma_1 \vert \Gamma_2\rangle
= (p+1)\langle \Gamma_1 \vert \Gamma_2\rangle$. So, $\forall\Gamma_1\in\mathcal{I}^{p+1}[D_1^{n_1} ]$, $\forall\Gamma_2\in\mathcal{I}^{p+1}[D_2^{n_2} ]$, $\langle \Gamma_1 \vert \Gamma_2\rangle=0$, which proves that, $D_1^{n_1}$ and $D_2^{n_2}$ are $(p+1)$-orthogonal.
The proof works for bosons, if $\wedge$ is replaced by $\vee$.
By induction, the result holds for all $q$ such that, $inf(n_1,n_2)\geq q\geq p$. 

So, $p$-orthogonality provides us with a graded orthogonality concept for states of identical particles, and the traditional term of ``strong orthogonality'' attached to $1$-orthogonality is justified in the sense that it implies $p$-orthogonality for all $p$. 

Example 1:
For integers, $n>p>0$, the pairs,  $\Psi_1:=\phi_1\wedge\ldots \wedge\phi_{n}$ and $\Psi_2:=\phi_{1}\wedge\ldots \wedge\phi_{n-p}\wedge\phi_{n+1}\ldots \wedge\phi_{n+p}$, (equivalent to those denoted with second quantization operators in the introduction), are $(n-p+1)$-orthogonal but not $(n-p)$-orthogonal since for $\Phi_1:=\phi_{n-p+1}\wedge\ldots \wedge\phi_{n}$ and $\Phi_2:=\phi_{n+1}\wedge\ldots \wedge\phi_{n+p}$, $\langle \Phi_1\hookleftarrow\Psi_1 \vert \Phi_2\hookleftarrow\Psi_2\rangle=\langle \phi_1\wedge\ldots \wedge\phi_{n-p} \vert \phi_1\wedge\ldots \wedge\phi_{n-p}\rangle=1$ is non zero, although $\left( \Phi_i\hookleftarrow\Psi_i\right)\in\mathcal{I}^{n-p}[\Psi_i ]$, for $i\in\{1,2\}$, by definition.\\
Example 2:
Let $(\phi_i)_{i=1, \ldots,8}$ be $8$ orthogonal spinorbitals.
The functions $\Psi_1:=\phi_1\wedge\phi_2\wedge\phi_3 + \phi_4\wedge\phi_5\wedge\phi_6$ and $\Psi_2:=\phi_1\wedge\phi_7 + \phi_2\wedge\phi_8$ are $2$-orthogonal (it is impossible to obtain $\Psi_2$ by annihilating a spinorbital in $\Psi_1$) but not $1$-orthogonal since both $\phi_1$ and $\phi_2$ belongs to their one-internal space.

\subsection{Araki angles}
Within a given ``graduation'', e.g. the set of functions which are $(p+1)$-orthogonal but not $p$-orthogonal for some $p$, a finer measure of non $p$-orthogonality is given by the Araki angles between the $p$-internal spaces.
The Araki angles between the spin $\alpha$- and the spin $\beta$-part of the one-internal spaces have already been introduced by the present author to study spin contamination in spin-unrestricted wave functions \cite{Cassam93-ijqc}. 
The cosines of these angles are the overlaps between biorthogonal functions.

Consider the $p$-internal spaces $\mathcal{I}^p[D_1^{n_1} ]$ and $\mathcal{I}^p[D_2^{n_2} ]$ of two density operators (or wave functions in case of pure states), with $n_1\leq n_2$.
Let us set $E:=\mathcal{I}^p[D_1^{n_1} ]+\mathcal{I}^p[D_2^{n_2} ]$ and denote $P_j\ (j\in\{1,2\})$ the orthogonal projector on $\mathcal{I}^p[D_j^{n_j} ]$ in $E$.
The construction is the same as that of \cite{Cassam93-ijqc}. We define the operators, ``$COS\Theta^p$'' and ``$SIN\Theta^p$'',
\begin{equation}
COS\Theta^p := \vert P_1 + P_2 - Id_E\vert,\ SIN\Theta^p := \vert P_1 - P_2\vert,
\end{equation} 
which satisfy,
\begin{equation}
(COS\Theta^p)^2 + (SIN\Theta^p)^2 =  Id_E.
\end{equation}
$(COS\Theta^p)^2$ is a Hermitian, positive operator, whose eigenvalues are in the interval $[0,1]$. One can associate to each eigenvalue, $\lambda^p_i$, an angle by,
\begin{equation}
\theta^p_i =  arccos(\sqrt{\lambda^p_i}).
\end{equation}
The eigenspaces of $(COS\Theta^p)^2$, that we write $V_{\theta^p_i}$ (rather than $V_{\lambda^p_i}$) decomposes $E$
into a direct sum of orthogonal vector subspaces, 
\begin{equation}
E:=\bigoplus_{\theta^p_i} V_{\theta^p_i}. 
\label{araki_decomp}
\end{equation}
The Araki angle operator, $\Theta^p$, is defined on $E$ as 
\begin{equation}
\Theta^p :=  \sum_i\ \theta^p_i\ .\ P_{V_{\theta^p_i}},
\end{equation}
where $ P_{V_{\theta^p_i}}$ is the orthogonal projector on $V_{\theta^p_i}$.
The remarkable property of the decomposition (\ref{araki_decomp}) is that it   ``respects'' the structure of the  $p$-internal spaces $\mathcal{I}^p[D_1^{n_1} ]$ and $\mathcal{I}^p[D_2^{n_2} ]$, in the sense that, for $j\in\{1,2\}$,
\begin{equation}
\mathcal{I}^p[D_j^{n_j} ]=\bigoplus_{\theta^p_i} \mathcal{I}^p[D_j^{n_j} ]\bigcap V_{\theta^p_i}. 
\label{araki_decomp2}
\end{equation}
Setting $\mathcal{I}^p[D_j^{n_j} ]_{\theta^p_i}:= \mathcal{I}^p[D_j^{n_j} ]\bigcap V_{\theta^p_i}$ we obviously have that $\mathcal{I}^p[D_1^{n_1} ]_{\theta^p_i}$ is orthogonal to $\mathcal{I}^p[D_2^{n_2} ]_{\theta^p_j}$ if $i\neq j$, and if $i= j$, any pair of function $\Phi_1\in\mathcal{I}^p[D_1^{n_1} ]_{\theta^p_i},\ \Phi_2\in\mathcal{I}^p[D_2^{n_2} ]_{\theta^p_i}$ can be thought geometrically as making an angle $\theta^p_i$. 

Particular cases:\\
If there is only one eigenvalue, $\lambda^p_1=0$, hence $\theta^p_1=\frac{\pi}{2}$, the states are in fact  $p$-orthogonal.\\
If the eigenvalue, $\lambda^p_i=1$, hence $\theta^p_i=0$, is present and its multiplicity equal to $n_1$, then the $p$--internal space $\mathcal{I}^p[D_1^{n_1} ]$ is a vector subspace of $\mathcal{I}^p[D_2^{n_2} ]$.

Between these extreme cases, the Araki angles provide us with a quantitative
mean to assess departure from $p$-orthogonality.

\section{Application to group functions}

When the respective states of two groups of identical particles are $p$-orthogonal, at most $p-1$ particles are possibly ``overlapping'' over the two groups, in the sense that the overlap of any $p$-particle  state occupied in the $n_1$-particle state of the first group with any $p$-particle  state occupied in the $n_2$-particle state of the other group is zero. In particular, for $p=1$, no particle overlaps and the two groups are distinguishable.

Let us emphasize that this notion of distinguishability does not necessarily imply the localization of the two groups of particles in two non-intersecting regions of real space. It has only to do with the orthogonality of abstract Hilbert spaces. 

A direct consequence of $p$-orthogonality is the cancellation of matrix elements between Hermitian operators that only couple a limited number of particles:\\
\textbf{Proposition 2:} 
Two $p$-orthogonal, $n$-particle states cannot be coupled through a $q$-particle interaction operator, $V^q$, if $q\leq n-p$.\\
\textbf{Proof:} 
A $q$-particle operator, $V^q$, is an operator that can be expressed in the second quantization formalism as,
\begin{equation}
V^q=\sum\limits_{\begin{array}[t]{c} ^{I:=(i_1,\ldots ,i_q),}\\[-20pt]  ^{J:=(j_1,\ldots ,j_q)}\end{array} } \lambda_{I,J}\ a^{\dagger}_{i_1}\ldots a^{\dagger}_{i_q}a_{j_q}\ldots a_{j_1}\ ,
 \label{q-particle}
\end{equation}
with $\lambda_{I,J}=\lambda_{J,I}^*$.
Let $\Psi_1$ and $\Psi_2$ be two $p$-orthogonal, $n$-particle wave functions. By linearity,\\
$\langle\Psi_1\vert V^q\vert\Psi_2\rangle = \sum\limits_{I, J}\lambda_{I,J}\ \langle\Psi_1\vert a^{\dagger}_{i_1}\ldots a^{\dagger}_{i_q}a_{j_q}\ldots a_{j_1}\vert\Psi_2\rangle = \sum\limits_{I, J}\lambda_{I,J}\ \langle a_{i_q}\ldots a_{i_1}\Psi_1\vert a_{j_q}\ldots a_{j_1}\Psi_2\rangle$. 
By definition, $\forall (i_1,\ldots ,i_q), (j_1,\ldots ,j_q), \quad a_{i_q}\ldots a_{i_1}\Psi_1\in\mathcal{I}^{n-q}[\Psi_1 ], \ a_{j_q}\ldots a_{j_1}\Psi_2\in\mathcal{I}^{n-q}[\Psi_2 ]$. By hypothesis, $p\leq (n-q)$, so Proposition 1 shows that all these pairs of $(n-q)$-particle wave functions are orthogonal:\\
$\forall (i_1,\ldots ,i_q), (j_1,\ldots ,j_q), \quad \langle a_{i_q},\ldots ,a_{i_1}\Psi_1\vert \ a_{j_q},\ldots ,a_{j_1}\Psi_2\rangle = 0$, hence  $\langle\Psi_1\vert V^q\vert\Psi_2\rangle = 0$.

In quantum chemistry, general antisymmetric product function methods \cite{Parr56,McWeeny59,Fock40,Hurley53,Lykos56,Kapuy58,Kapuy59a,Kapuy59b,Kapuy60a,Kapuy60b,Kapuy61a,Kapuy61b,Kapuy62,Kroner60,Klessinger65,Klessinger70,McWeeny60,McWeeny63,Angyan00,Li02,Rassolov02,Rosta00,Rosta02,Tokmachev05-jcc,Tokmachev06-ijqc,Tokmachev06-cp,Cassam06-jcp} optimize $n$-electron wave functions of the form:

\begin{equation}
\Psi=\Psi_1\wedge\cdots\wedge\Psi_r \ , 
\label{groupF}
\end{equation} 

where $\Psi_i$ is an $n_i$-electron function and $\sum_i n_i =n$. So far, for practical purposes, all these approaches (except those of \cite{McWeeny63,Cassam06-jcp}), have imposed the constraint that the $\Psi_i$'s have to be  $1$-orthogonal to one another. In the Electronic Mean Field Configuration Interaction (EMFCI) approach \cite{Cassam06-jcp},
no orthogonality constraint is \textit{a priori} imposed on the $\Psi_i$. In particular, in the simple case where, for all $i$, $n_i=2$, both APSG (Antisymmetrized Product of Strongly orthogonal Geminals) \cite{Rosta00} and  AGP (Antisymmetrized Geminal Product) of extreme type \cite{Coleman00}, $\Psi_1\wedge\cdots\wedge\Psi_1$, are considered by the EMFCI variational process.  

Therefore, it would be interesting to analyse the optimized EMFCI functions obtained for different systems and geometries in terms of their $p$-orthogonality properties, and see for instance, if they are closer to the APSG case ($1$-orthogonality) or to the AGP of extreme type case (non $2$-orthogonality with the Araki angle equal to zero for all pairs of two-electron group functions).
However, in the present study we will limit ourselves to emphasize how enforcing a $p$-orthogonality constraint between the $\Psi_i$'s, simplifies the computation of the Hamiltonian and overlap matrix elements between general antisymmetric product functions (Eq.(\ref{groupF})). 

Let us consider another such function, $\Psi'=\Psi_1'\wedge\cdots\wedge\Psi_r'$, (with $n_i'=n_i$), and define the notation, $$\Psi_{\hat{i}}:=\Psi_1\wedge\cdots\wedge\Psi_{i-1}\wedge\Psi_{i+1}\wedge\cdots\wedge\Psi_r,\qquad    \Psi_{\hat{i}\hat{j}}:=(\Psi_{\hat{i}})_{\hat{j}},\qquad and\ so\ on,$$  to denote that one or more specified factors have been taken out of a product function. So, for example $\Psi'=\Psi_1'\wedge\Psi_{\hat{1}}'$. It can be shown, using the Hopf algebra tools of \cite{Cassam06-jcp} 
, that,


\begin{eqnarray}
\lefteqn{\langle \Psi_1'\wedge\Psi_{\hat{1}}'\vert \Psi_1\wedge\cdots\wedge\Psi_r\rangle = }\nonumber\\[10pt]
\lefteqn{\sum\limits_{\begin{array}[t]{c} ^{I^1,\ldots , I^r}\\[-20pt]  ^{\forall j\ \vert I^j\vert\in\{0,\ldots ,  n_j\},}\\[-20pt] ^{\sum_j\ \vert I^j\vert=n_1 }\end{array}}
\rho_{\vert I^1\vert,n_1-\vert I^1\vert,\ldots,\vert I^r\vert,n_r-\vert I^r\vert}\ \langle \Psi_1'\vert  
(\Psi_1)_{I^1}\wedge\cdots\wedge(\Psi_r)_{I^r}\rangle \langle \Psi_{\hat{1}}'\vert  (\Psi_1)_{\bar{I}^1}\wedge\cdots\wedge(\Psi_r)_{\bar{I}^r}\rangle}\nonumber\\
\label{overlap}
\end{eqnarray} 
where \footnotemark[1],
\footnotetext[1]{this formula is a particular case of a formula given in \cite{Cassam06-jcp,Cassam05_loutraki} with an error on the summation bounds. A correct version is, 
\begin{equation}
\rho_{n_1^1,\ldots ,n_p^1,\ldots ,n_1^q,\ldots ,n_p^q} = (-1)^{\sum\limits_{1\leq j<l\leq q}\ \sum\limits_{p\geq i>k\geq 1}  n_k^l \cdot n_i^j }
\label{rho_twist}
\end{equation}}
\begin{equation}
\rho_{\vert I^1\vert,n_1-\vert I^1\vert,\ldots,\vert I^r\vert,n_r-\vert I^r\vert} = (-1)^{\sum\limits_{j=2}^r  \sum\limits_{k=1}^{j-1} \vert I^j\vert \cdot (n_k- \vert I^k\vert)  },
\end{equation}
and where, for any $p$-particle wave function, $\Phi:=\sum\limits_{K:=(k_1<\ldots <k_p)}\lambda_{K}\ \psi_{k_1}\wedge\cdots\wedge \psi_{k_p}$, ($K$ runs over ordered sequences of positive integers, $(\psi_i)_i$ denotes a one-particle basis set), and any ordered sequence of length $m\in\{0,\ldots p\}$, $I:=(1\leq i_1<\ldots <i_m\leq p)$, the following compact notation is extensively used,
\begin{eqnarray}
\lefteqn{\cdots(\Phi)_{I}\cdots (\Phi)_{\bar{I}}\cdots:=\rho_{I,\bar{I}}\sum\limits_{K:=(k_1<\cdots <k_p)}\lambda_{K}\ \cdots(\psi_{k_{i_1}}\wedge\cdots\wedge \psi_{k_{i_m}}) }\nonumber\\
\lefteqn{\qquad\qquad\qquad\qquad\qquad\qquad\qquad\qquad\qquad\qquad\qquad\cdots (\psi_{k_{\bar{i}_1}}\wedge\cdots\wedge \psi_{k_{\bar{i}_{p-m}}})\cdots}
\end{eqnarray} 
with $\bar{I}:=(1\leq \bar{i}_1<\ldots <\bar{i}_{p-m}\leq p)$, complement of $I$ in $\{1<2<\cdots < p \}$, $\rho_{I,\bar{I}}$ is the sign of the permutation reordering the concatenated sequence $I//\bar{I}$ in increasing order; if the length, $\vert I\vert$, of $I$ is $0$ then, by convention, $(\Phi)_{I}:=(\Phi)_{\emptyset}=1$, and $\rho_{\emptyset,(1<\cdots <p)}=1$; note that, $(\Phi)_{(1<\cdots <p)}=\Phi$.

If we assume that the group-$1$ (called the group of active electrons in the EMFCI method) wave function, $\Psi_1'$, is $q$-orthogonal to the product of the wave functions of the other groups (called spectator groups in the EMFCI method), $\Psi_{\hat{1}}=\Psi_2\wedge\cdots\wedge\Psi_r$, Eq.(\ref{overlap}) becomes,
\begin{eqnarray}
\lefteqn{\langle \Psi_1'\wedge\Psi_{\hat{1}}'\vert \Psi_1\wedge\cdots\wedge\Psi_r\rangle = }\nonumber\\[10pt]
\lefteqn{\sum\limits_{\begin{array}[t]{c} ^{I^1,\ldots , I^r}\\[-20pt]  ^{\vert I^1\vert\in\{n_1-q+1,\ldots ,  n_1\},}\\[-20pt]  ^{\forall j>1\ \vert I^j\vert\in\{0,\ldots ,  n_j\},}\\[-10pt] ^{\sum\limits_{j=1}^r\ \vert I^j\vert=n_1} \end{array}}
\rho_{\vert I^1\vert,n_1-\vert I^1\vert,\ldots,\vert I^r\vert,n_r-\vert I^r\vert}\ \langle \Psi_1'\vert  
(\Psi_1)_{I^1}\wedge\cdots\wedge(\Psi_r)_{I^r}\rangle \langle \Psi_{\hat{1}}'\vert  (\Psi_1)_{\bar{I}^1}\wedge\cdots\wedge(\Psi_r)_{\bar{I}^r}\rangle}\nonumber\\
\label{overlap-q}
\end{eqnarray} 
that is to say, the summation on the ordered sequences, $I^1$, is limited to those with length strictly more than $n_1-q$. Without this restriction, the number of $I^1$-sequences would be ${n \choose n_1}=\sum\limits_{i=0}^{n_1}{n_1 \choose i}{n-n_1 \choose n_1-i}$, whereas with the $q$-orthogonality restriction, it falls down to $\sum\limits_{i=n_1-q+1}^{n_1}{n_1 \choose i}{n-n_1 \choose n_1-i}$. In the limit case of $1$-orthogonality, only the sequence $I^1=(1<2<\cdots <n_1)$ remains, and Eq.(\ref{overlap-q}) simplifies to, 
\begin{equation}
\langle \Psi_1'\wedge\Psi_{\hat{1}}'\vert \Psi_1\wedge\cdots\wedge\Psi_r\rangle = \langle \Psi_1'\vert  
\Psi_1\rangle \langle \Psi_{\hat{1}}'\vert  \Psi_2\wedge\cdots\wedge\Psi_r\rangle.
\label{overlap-1}
\end{equation} 
At the other end, enforcing $n_1$-orthogonality between active and spectator groups, that is the weakest $q$-orthogonality constraint, rules out only the case $I^1=\emptyset$. However, this eliminates already ${n-n_1 \choose n_1}$ $I^1$-sequences, and this number becomes comparable to the total number of $I^1$-sequences, ${n \choose n_1}$, in the limit of practical interest where the number of active electrons, $n_1$, is small with respect to the total number of electrons, $n$. 

Consider now the matrix elements between general antisymmetric product functions, of an operator, $H$, whose action on $n$-electron wave functions is induced by a $s$-particle operator, $\hat{h}$, (with $s\leq n$). Typically, $\hat{h}$ will be a Coulombian Hamiltonian, so that $s=2$. Its induced action on the $n$-electron wave function of Eq.(\ref{groupF}) can be expressed using Hopf algebra techniques as,
\begin{eqnarray}
\lefteqn{ H[\Psi_1\wedge\cdots\wedge\Psi_r]=}\nonumber\\[10pt]
&\sum\limits_{\begin{array}[t]{c} ^{J^1,\ldots , J^r}\\[-20pt]  ^{\forall j\ \vert J^j\vert\in\{0,\ldots ,  n_j\},}\\[-20pt] ^{\sum_j\ \vert J^j\vert=n-s }\end{array}}
\rho_{\vert J^1\vert,\vert \bar{J^1}\vert,\ldots,\vert J^r\vert,\vert \bar{J}^r\vert}
(\Psi_1)_{J^1}\wedge\cdots\wedge(\Psi_r)_{J^r}\wedge \hat{h}\left[ (\Psi_1)_{\bar{J}^1}\wedge\cdots\wedge(\Psi_r)_{\bar{J}^r}\right].&
\label{h_action}
\end{eqnarray} 
So, a matrix element, $\langle  \Psi_1'\wedge\Psi_{\hat{1}}'\vert H[\Psi_1\wedge\cdots\wedge\Psi_r]\rangle$, is a sum of terms of the form,  $\langle  \Psi_1'\wedge\Psi_{\hat{1}}'\vert (\Psi_1)_{J^1}\wedge\cdots\wedge(\Psi_r)_{J^r}\wedge \hat{h}\left[ (\Psi_1)_{\bar{J}^1}\wedge\cdots\wedge(\Psi_r)_{\bar{J}^r}\right]\rangle$, similar to Eq.(\ref{overlap}) but with, in the ket, $(r+1)$ groups of $\vert J^1\vert,\ldots,\vert J^r\vert,s$ particles respectively, instead of $r$ groups of $n_1,\ldots ,n_r$ particles. In particular, $\vert J^1\vert$ can be less than $n_1$, the number of particles in $\Psi_1'$. However, essentially the same development can be carried out, the $q$-orthogonality constraint limiting the summation for each term,
\newpage
\begin{eqnarray}
\lefteqn{\langle\Psi_1'\wedge\Psi_{\hat{1}}'\vert (\Psi_1)_{J^1}\wedge\cdots\wedge(\Psi_r)_{J^r}\wedge \hat{h}\left[(\Psi_1)_{\bar{J}^1}\wedge\cdots\wedge(\Psi_r)_{\bar{J}^r}\right]\rangle = }\nonumber\\[10pt]
\lefteqn{ \sum\limits_{\begin{array}[t]{c} ^{I^1,\ldots , I^{r+1}}\\[-20pt] ^{\vert I^1\vert\in\{\vert J^1\vert-q+1,\ldots ,  \vert J^1\vert\}}\\[-20pt] ^{1<j\leq r,\ \vert I^j\vert\in\{0,\ldots , \vert J^j\vert\}}\\[-20pt] ^{\vert I^{r+1}\vert\in\{0,\ldots ,  s\},}\\[-10pt] ^{\sum\limits_{j=1}^{r+1}\ \vert I^j\vert=n_1 }\end{array}}\rho_{\vert I^1\vert,\vert \bar{I^1}\vert,\ldots,\vert I^{r+1}\vert,\vert \bar{I}^{r+1}\vert}}\nonumber\\[-90pt]
\lefteqn{\qquad\qquad\qquad\qquad\qquad\cdot\ \langle \Psi_1'\vert  
\left( (\Psi_1)_{J^1}\right)_{I^1}\wedge\cdots\wedge\left( (\Psi_r)_{J^r}\right)_{I^r}\wedge \left( \hat{h}\left[ (\Psi_1)_{\bar{J}^1})\wedge\cdots\wedge(\Psi_r)_{\bar{J}^r}\right]\right)_{I^{r+1}}\rangle }\nonumber\\[10pt]
\lefteqn{\qquad\qquad\qquad\qquad\qquad\cdot\ \langle \Psi_{\hat{1}}'\vert  \left( (\Psi_1)_{J^1}\right)_{\bar{I}^1}\wedge\cdots\wedge\left( (\Psi_r)_{J^r}\right)_{\bar{I}^r}\wedge \left( \hat{h}\left[ (\Psi_1)_{\bar{J}^1})\wedge\cdots\wedge(\Psi_r)_{\bar{J}^r}\right]\right)_{\bar{I}^{r+1}}\rangle.}\nonumber\\
\label{h-el-q}
\end{eqnarray}

\section{Conclusion and prospects}
 
We have defined the geometric concept of $p$-orthogonality between quantum states of sets of identical particles.
This concept provides us with a graded measure of indistinguishability in the sense that two sets of identical particles that are $q$-orthogonal can be seen as ``more indistinguishable'' 
than two sets that are $p$-orthogonal if $q>p$,
because a larger subset of particles can possibly share i.e. occupy a substate, common i.e. internal to the  quantum states of both sets. 

A classical anology can be attempted with the case of two groups of  billiard balls of the same color.
When $p=1$, no particle is mixed
and the two sets of particles are in fact distinguishable like two sets of balls localized in distinct areas of a billiard  table. Pushing this classical picture one step beyond for $p>1$, the two sets of balls would be connected but at most $p-1$ balls of one set would be in contact with at most $p-1$ balls of the other set. So, the smaller $p$, the narrower the bridge between the two sets of balls would be. Assuming that the group of origin of the balls making up the bridge is unknown, these balls would be the analogues of the genuinely indistinguishable particles which belong partially to both sets. 

Let us emphasize that this classical picture should not be carried too far, for, in particular, our notion of distinguishability does not imply the localization of the two groups of particles in two non-overlapping regions of real space. It has only to do with the orthogonality of Hilbert spaces called the $p$-internal spaces of the quantum states. $p$-orthogonality can be seen as a mathematical and quantum mechanical rigorous formalisation of this classical  image.

$p$-orthogonality can be used to remove some arbitrariness in the choice of a representation for a quantum system in the same manner as localization criteria do. For example, consider $n$ pairs of spin-$\frac{1}{2}$ fermions
whose state can be represented by a Slater determinant, $\Psi:=\phi_1\wedge\bar{\phi}_1\wedge\cdots \wedge\phi_{n}\wedge\bar{\phi}_{n}$, where $\phi_1, \ldots ,\phi_{n}$, are orthonormal one-fermion functions of spin $z$-component $\frac{1}{2}$, and  $\bar{\phi}_1,\ldots , \bar{\phi}_{n}$, their counterparts with spin $z$-component equal to $- \frac{1}{2}$ . Such a wave function is invariant within a phase factor under an unitary transformation, $u$, of the one-particle functions, $\phi_1,\ldots ,\phi_{n}$. There are various techniques \cite{Foster60,Edmiston63} that exploit this freedom to reexpress a wave function with a new set of one-particle functions, $\psi_j=u(\phi_j)$, localized in real space, and such that $\Psi$ has still the form of a Slater determinant, $\Psi:=\psi_1\wedge\bar{\psi}_1\wedge\cdots \wedge\psi_{n}\wedge\bar{\psi}_{n}$. However, this only provides a constraint on one-particle states and there is still more freedom available. For example,  $\Psi$ can be re-expressed as an AGP of extreme type with the same localized one-particle functions, 
\begin{equation}
\Psi=\underbrace{g\wedge g\wedge\cdots\wedge g}_{n factors},
\label{HF2}
\end{equation}
where $g=(n!)^{-\frac{1}{n}}(\psi_1\wedge \bar{\psi}_1+\cdots + \psi_{n}\wedge\bar{\psi}_{n})$. If we set, $g_i=\psi_i\wedge \bar{\psi}_i$ for all $i$, we also have 
\begin{equation}
\Psi=g_1\wedge g_2\wedge\cdots\wedge g_n.
\label{HF1}
\end{equation}
Imposing  $1$-orthogonality or even $2$-orthogonality between the two-fermion functions appearing in Eqs.(\ref{HF2}) and (\ref{HF1}) can discriminate between these two equivalent writings.

The graded structure of $p$-orthogonality constraints naturally leads one to consider a corresponding hierarchy of approximations for methods based on general antisymmetric product functions.
In this work, we have exhibited the link between $p$-orthogonality and the combinatorics involved in the calculation of the matrix elements of particle-number-preserving observables. In the frame of the EMFCI method, we have shown that even the weakest 
$p$-orthogonality constraint between an active group of particles and the rest of spectator particles can be effective in limiting the computational effort required for the calculation of matrix elements. We will report shortly on the accuracy of EMFCI wave functions constrained by $p$-orthogonality, for increasing value of $p$ on a benchmark of molecular systems.

\section*{Acknowledgements}
We acknowledge the french ANR for fundings (Project AHBE). We are indebted to Dr. F. Patras for Eq.(\ref{rho_twist}) and for helpful discussions, which have contributed to the maturation of this work.

\end{document}